\documentclass[12pt]{article}
\usepackage{amsfonts}
\usepackage{amssymb}
\usepackage{amsbsy}
\usepackage{graphics}

\input epsf.sty

\newlength{\TZ}
\newcommand{\BEQ}{\begin{equation}}     % Gleichungen Anfang ..
\newcommand{\BEA}{\begin{eqnarray}}
\newcommand{\EEQ}{\end{equation}}       % .. und Ende
\newcommand{\EEA}{\end{eqnarray}}
\def\be{\begin{equation}}
\def\ee{\end{equation}}
\def\ba{\begin{eqnarray}}
\def\ea{\end{eqnarray}}
\newcommand{\D}{{\rm d}}                % gerades d fuer Ableitungen
\newcommand{\II}{{\rm i}}               % gerades i fuer komplexe Einheit
\newcommand{\wit}[1]{\widetilde{#1}}    % weite Schlange
\newcommand{\lap}[1]{\overline{#1}}     % Querstrich oben
\renewcommand{\vec}[1]{\boldsymbol{#1}} % Vektoren fettgedruckt

\newcommand{\zeile}[1]{\vskip #1 \baselineskip} % N Zeilen ueberschlagen
\catcode`\@=11
\def\numberbysection{\@addtoreset{equation}{section}
        \def\theequation{\thesection.\arabic{equation}}}
\numberbysection

\begin{document}

\begin{titlepage}

~~~ 
%%{\hfill \tt \today; pr\'eliminaire !}

\vskip 1.5 cm
\begin{center}
{\Large \bf Competition between dynamic and thermal relaxation in
non-equilibrium spin systems above the critical point}
\end{center}

\vskip 2.0 cm
\centerline{  {\bf Alan Picone}$^a$, {\bf Malte Henkel}$^{a,c}$ and
{\bf Jean Richert}$^b$}
\vskip 0.5 cm
\centerline {$^a$Laboratoire de Physique des
Mat\'eriaux,\footnote{Laboratoire associ\'e au CNRS UMR 7556}
Universit\'e Henri Poincar\'e Nancy I,}
\centerline{ B.P. 239,
F -- 54506 Vand{\oe}uvre l\`es Nancy Cedex, France}
\vskip 0.5 cm
\centerline{$^b$Laboratoire de Physique
Th\'eorique,\footnote{Laboratoire associ\'e au CNRS UMR 7085}
Universit\'e Louis Pasteur,}
\centerline{3, rue de l'Universit\'e, F -- 67084 Strasbourg Cedex, France}
\vskip 0.5 cm
\centerline{$^c$Centro de F\'{\i}sica da Mat\'eria Condensada,
Universidade de Lisboa,}
\centerline{Av. Prof Gama Pinto 2, P -- 1649 - 003 Lisboa, Portugal}

\begin{abstract}

We study the long-time behaviour and the spatial correlations of a simple
ferromagnetic spin system whose kinetics is governed by a thermal bath
with a time-dependent temperature which is characterized by a given
external relaxation time $\tau$. Exact results are obtained
in the framework of the spherical model in $d$ dimensions.
In the paramagnetic phase, the long-time kinetics is shown to depend
crucially on the ratio between $\tau$ and the internal equilibration
time $\tau_{\rm eq}$.

If $\tau\lesssim \tau_{\rm eq}$, the model relaxes rapidly towards an 
equilibrium state but there appear transient and spatially oscillating 
contributions in
the spin-spin correlation function. On the other hand, if
$\tau\gg \tau_{\rm eq}$ the system is clamped and its time evolution may be 
delayed with respect to the one of the heat bath. 
For waiting times $s$ such that $\tau\gg s\gg \tau_{\rm eq}$, 
a quasi-stationary state is found where
the fluctuation-dissipation theorem does not hold. 
\end{abstract}

\zeile{4}
\noindent
PACS numbers: 05.70.Ln, 05.70.Jk, 05.40+j, 82.20.M 
\end{titlepage}

%%%%%%%%%%%%%%%%%%%%%%%%%%%%%%%%%%%%%%%%%%%%%%%%%%%%%%%%%%%%%%%%%%%%%%%%%%%%%%%%
\section{Introduction}
%%%%%%%%%%%%%%%%%%%%%%%%%%%%%%%%%%%%%%%%%%%%%%%%%%%%%%%%%%%%%%%%%%%%%%%%%%%%%%%%

Non-equilibrium systems generally display specific features which are absent
from their corresponding thermodynamically equilibrated counterparts. For
example, consider a many-body system contained in a thermostat and prepared in
some initial state. Then assume that the temperature of the heat bath is 
rapidly quenched to a different temperature $T_1$ and the system is then 
allowed to evolve freely, with the temperature fixed at $T_1$.
In this setting, the system under study may show ageing effects, which come
from the dynamical breaking of time-translation invariance,
see \cite{Stru78,Bray94,Bouc98,Cate00} for reviews.
Although ageing is known, since a very long time, to occur in glassy systems, 
similar effects have more recently also been observed for
non-disordered spin systems. In simple ferromagnets such as the Ising model,
ageing effects have been studied through the long-time behaviour of two-time
correlation and response functions, see \cite{Godr02} and references therein.
In addition, it has been proposed recently that the well-known dynamical
scale invariance, associated with the ageing phenomenon
(see e.g. \cite{Stru78,Bray94}) in simple
ferromagnets, may be generalized towards a larger group of local dynamical
scale transformations \cite{Henk02}. Among other results this theory
leads to an explicit prediction for the time-dependent thermoremanent
magnetization which has been confirmed in
several distinct models \cite{Henk01,Pico02,Cala02,Cann01,Cala02a}.

However, current theoretical studies generally assume that the quench
is infinitely rapid, or in other words that the
final temperature $T_1$ is reached immediately. In real experiments, this
condition is often far from being satisfied, see e.g. \cite{Grig99,Krue02}. 
It is therefore interesting to
investigate the behaviour of a system coupled to a bath with a time-dependent
temperature $T(t)$. Here, we shall study the long-time behaviour of a system
initially prepared in a fully disordered state and brought at time $t=0$ into 
contact with a thermal bath which evolves from an initial temperature
$T_0$ to a final temperature $T_1$ in a finite or infinite interval of
time. In the present paper, we shall study the
following form of a time-dependent temperature
\BEQ \label{1:eqTt}
T(t) = T_1 + (T_0 - T_1) \exp\left( - t/\tau\right)
\EEQ
where $\tau$ is a given relaxation time which sets an external time scale
$\tau_{\rm ext}=\tau$. Certainly, the long-time behaviour of a statistical
system will depend, besides on the external time scale $\tau$, also on the
internal time scale(s) $\tau_{\rm int}(T)$ which arise from the internal
self-organization of the system, governed by the microscopic
interactions. Two basic cases should be expected. First, if
$\tau_{\rm int}\gg \tau$, the system simply relaxes towards equilibrium
at the final temperature $T_1$. On the hand, if $\tau_{\rm int}\ll \tau$,
the system is said to be {\em clamped\/} at the external temperature $T(t)$
and its time history will depend on the properties of the external heat bath.
While for times $t\gg\tau\gg\tau_{\rm int}$ the system should have simply 
relaxed back to equilibrium, 
new effects might be expected to occur in the regime 
$\tau\gg t\gg\tau_{\rm int}$ and this may allow to distinguish
between these two cases. The
time-dependence as specified in eq.~(\ref{1:eqTt}) should contain the basic
ingredients which are needed for a conceptual understanding of this kind
of phenomenon.\footnote{Originally, we became interested in this problem 
through the consideration of the fragmentation of microscopic
systems like atomic nuclei where it has often been assumed that the detected
fragments are at thermodynamic equilibrium and fixed temperature \cite{Rich01}.
However, such descriptions do not take care of the time evolution of these
systems which expand in space over a finite time interval and cool down before
they are detected.}

One of the central questions is whether/when under the conditions just
described the system may evolve towards thermodynamic equilibrium. A
convenient way to measure the distance of a system from equilibrium is
through the fluctuation-dissipation ratio \cite{Cugl94a,Cugl94b}
\BEQ
X(t,s) = T R(t,s) \left( \frac{\partial C(t,s)}{\partial s}\right)^{-1}
\EEQ
where $C(t,s)$ and $R(t,s)$ are the two-time auto-correlation and autoresponse
functions, respectively (see section 2 for the precise definitions).
At equilibrium, the fluctuation-dissipation theorem states that $X(t,s)=1$.
Therefore, it is an important question under which physical conditions
equilibrium can be reached. This question has received much attention
in glasses, see \cite{Garr01,Pere02} for recent theoretical and 
\cite{Grig99,Heri02,Bell02} for recent experimental examples.
Understanding this point must come before the more complicated questions
relating to an eventual ageing behaviour can be addressed.

In order to obtain explicit analytical results, we shall study the
exactly solvable kinetic spherical model with a time-dependent bath
temperature, to be defined precisely in section 2.
For constant temperatures, this model has been studied in great detail in the
past, either in the context of continuum field theories
\cite{Bray91,Jans89,Newm90,Kiss93,Coni94,Cala02} or else in the form
of a lattice model \cite{Cugl95,Zipp00,Godr00b,Cann01,Corb02,Pico02,Fusc02}. 
It is well-established that the spherical model, 
despite its technical simplicity,
still contains the main physical features encountered in other systems
(such as the Ising model) which might be considered to be closer to the
experimental reality but which no longer permit an exact analytical solution.
In particular, results found from the spherical model in $d<4$ space dimensions
are known to be different from the predictions of mean-field theory.
In this paper, we shall be concerned exclusively with the question under what
condition a stationary or quasi-stationary state of the model 
may be considered to be an equilibrium state in the sense that $X(t,s)=1$. 
This question can be studied by restricting attention to the high-temperature 
phase such that the system is brought into contact with a heat bath
at the initial temperature $T_0$. The heat bath then cools down to 
the final temperature $T_1$ such that both
are above the critical temperature, viz. $T_0>T_1>T_c>0$ 
and the model remains in the disordered high-temperature phase. 
In that phase, there is no ageing behaviour
to be expected. We leave the question of a possible ageing behaviour in
systems coupled to a time-dependent bath for future work.

The content of the paper is as follows. In section 2 we introduce the model
and obtain the formal exact solution. The equal-time correlator is analysed
in section 3 and we discuss several new non-trivial time and length scales
which arise. Two-time quantities and the fluctuation-dissipation ratio are
studied in section 4 and the results
are generalized to include the clamped case in section 5. Section 6
presents our conclusions.

%%%%%%%%%%%%%%%%%%%%%%%%%%%%%%%%%%%%%%%%%%%%%%%%%%%%%%%%%%%%%%%%%%%%%%%%%%%%%%%%
\section{Model and formalism}
%%%%%%%%%%%%%%%%%%%%%%%%%%%%%%%%%%%%%%%%%%%%%%%%%%%%%%%%%%%%%%%%%%%%%%%%%%%%%%%%

We begin by recalling the definition of the kinetic spherical model,
using the formalism as exposed in \cite{Cugl95,Godr00b,Pico02}.
We consider a system of time-dependent classical
spin variables $S_{\vec{x}}(t)$ located on the sites $\vec{x}$ of
a $d$-dimensional hypercubic lattice. They may take arbitrary real values
subject only to the spherical constraint
\BEQ
\sum_{\vec{x}} S_{\vec{x}}(t)^2 = {\cal N}
\label{eq0}
\EEQ
where $\cal N$ is the number of sites of the lattice.
The spherical model Hamiltonian reads
\BEQ
{\cal H} = - J\sum_{<\vec{x},\vec{y}>}S_{\vec{x}}(t)S_{\vec{y}}(t)
\label{eq1}
\EEQ
where the sum extends over nearest-neighbour pairs only. In the following we
choose units such that $J=1$.
The system is supposed to be translation-invariant in all directions.
The kinetics is assumed to be described in terms of a Langevin equation 
\BEQ
\frac{\D S_{\vec{x}}(t)}{\D t} = \sum_{\vec{y}(\vec{x})}S_{\vec{y}}(t) 
-(2d+\mathfrak{z}(t))S_{\vec{x}}(t) + \eta_{\vec{x}}(t)
\label{eq2}
\EEQ
where the sum over $\vec{y}$ extends over the nearest neighbours of $\vec{x}$
and $\eta_{\vec{x}}(t)$ corresponds to a stochastic force which describes the
action of an environment which lies outside of the nearest-neighbour range.
Physically, this means that the model is assumed to be immersed into a
heat bath. The resulting forces are
supposed to be gaussian and thus to be characterized by an ensemble
average which is zero, viz. $\langle \eta_{\vec{x}}(t)\rangle =0$ and a second
moment which reads
\BEQ
\langle\eta_{\vec{x}}(t) \eta_{\vec{y}}(t')\rangle =
2 T(t)\,\delta_{\vec{x},\vec{y}}  \delta (t-t')
\label{eq3}
\EEQ
At time $t=0$ the system is brought into contact with the heat bath.
We assume in this work that the temperature of the heat bath decreases from
$T_0$ to $T_1$ with a characteristic relaxation time $\tau$ according to
\BEQ
T(t) = T_1 + (T_0 - T_1)\exp(-t/\tau)
\label{eq4}
\EEQ
The choice of this temporal behaviour is dictated by computational simplicity.
In addition, as we shall see, it already contains the basic effects which
will also be present for more general thermal histories.
For sufficiently large values of the initial bath temperature $T_0$, 
the typical correlation lengths are of the order of a lattice constant or less. 
Then for all practical purposes, the system is effectively uncorrelated.  
Finally, the function $\mathfrak{z}(t)$ is fixed by the spherical constraint
(\ref{eq0}) and has to be determined. In order to do this, 
we follow the standard procedure of replacing the spherical constraint 
by its mean value. It can be
shown that this does not change results if the infinite-system limit 
${\cal N}\to\infty$ is taken before the long-time limit 
$t\to\infty$ \cite{Fusc02}. 

By a Fourier transformation
\BEQ
\wit{f}(\vec{q}) = \sum_{\vec{r}} f_{\vec{r}} e^{-\II \vec{q}\cdot\vec{r}}
\;\; , \;\;
f_{\vec{r}} = (2\pi)^{-d} \int_{{\cal B}}\!\D\vec{q}\, \wit{f}(\vec{q})
e^{\II \vec{q}\cdot\vec{r}}
\EEQ
where the integral is taken over the first Brillouin zone $\cal B$, the
Fourier-transformed spin variable $\wit{S}(\vec{q},t)$ becomes
\BEQ
\wit{S}(\vec{q},t) = \frac {e^{-\omega(\vec{q})t}}{\sqrt{g(t)}}
\left[\wit{S}(\vec{q},0) +
\int_0^t\!\D t'\:
e^{\omega(\vec{q})t'} \sqrt{g(t')}~ \wit{\eta}(\vec{q},t')\right]
\label{eq6}
\EEQ
with the dispersion relation
\BEQ
\omega(\vec{q}) = 2\sum_{i=1}^d{\left(1-\cos(q_i)\right)}
\label{eq7}
\EEQ
and we have also defined
\BEQ
g(t)= \exp\left({2\int_0^t\!\D t'\: \mathfrak{z}(t')}\right)
\label{eq5}
\EEQ
Clearly, the time-dependence of $\wit{S}(\vec{q},t)$ and any correlators
will be given in terms of the function $g=g(t)$.

We now derive the expressions for the correlators and response functions for
an arbitrary thermal history $T=T(t)$. We begin with the equal-time spin-spin
correlation function
\BEQ
C_{\vec{x},\vec{y}}(t) =C_{\vec{x}-\vec{y}}(t) =
\langle S_{\vec{x}}(t) S_{\vec{y}}(t) \rangle
\label{eq8}
\EEQ
We obtain the Fourier transform $\wit{C}(\vec{q},t)$ from
\BEQ
\langle \wit{S}(\vec{q},t) \wit{S}(\vec{q}',t')\rangle =
{(2\pi)^d}\delta(\vec{q}+\vec{q}') \, \wit{C}(\vec{q},t)
\label{eq11}
\EEQ
and immediately find
\BEQ
\wit{C}(\vec{q},t) = \frac{e^{-2\omega(\vec{q})t}}{g(t)}
\left[{\wit{C}(\vec{q},0)} +
2\int_0^t\!\D t'\: T(t')\, e^{2\omega(\vec{q})t'}g(t')\right]
\label{eq12}
\EEQ
If we assume uncorrelated initial conditions\footnote{The treatment of
correlated initial conditions, following the lines of \cite{Pico02}, presents
no additional difficulty.}
\BEQ
C_{\vec{x},\vec{y}}(0) = \delta_{\vec{x}-\vec{y},\vec{0}}
\label{eq10}
\EEQ
as we shall always do in the following, we have $\wit{C}(\vec{q},0)=1$.
Because of the spherical constraint (\ref{eq1}) and spatial translation
invariance, the autocorrelator must satisfy
\BEQ
C_{\vec{0}}(t)= \int_{{\cal B}} \!\D \vec{q}\:
\wit{C}(\vec{q},t) = \langle S_{\vec{x}}(t)^2\rangle = 1
\label{eq9}
\EEQ
This in turn fixes $\mathfrak{z}(t)$ or via (\ref{eq5}) the function
$g(t)$ as the solution of a Volterra integral equation
\BEQ
g(t)= f(t) + 2{\int_0^t\!\D t'\: T(t')f(t-t')g(t')}
\label{eq15}
\EEQ
where
\BEQ
f(t) = \frac{1}{(2\pi)^d}\int_{{\cal B}}\!\D \vec{q}\: e^{-2\omega(\vec{q})t}
= \left( e^{-4t} I_0(4t)\right)^d
\label{eq16}
\EEQ
and $I_0$ is a modified Bessel function \cite{Abra65}. Once we have
found the solution of eq.~(\ref{eq15}), the correlation function can be
obtained.

Before doing this, we now give the expression for the two-time
correlation function $C_{\vec{x}-\vec{y}}(t,s)=
\langle S_{\vec{x}}(t)S_{\vec{y}}(s)\rangle$ and the two-time response function
$R_{\vec{x}}(t,s)$. The calculation follows entirely standard lines
\cite{Cugl95,Godr00b,Pico02} and we merely quote the result. In Fourier space
\BEQ \label{eqC}
\wit{C}(\vec{q},t,s) = \wit{C}(\vec{q},s) e^{-\omega(\vec{q})(t-s)}
\sqrt{\frac{g(s)}{g(t)}}
\EEQ
and where $g=g(t)$ is the solution of eq.~(\ref{eq15}).
Similarly, the response function is obtained in the usual way
\cite{Newm90,Kiss93,Cugl95,Godr00b,Pico02}
by adding a small magnetic
field term $\delta{\cal H} = - \sum_{\vec{x}} h_{\vec{x}}(t) S_{\vec{x}}(t)$
to the Hamiltonian. We easily find in Fourier space
\BEQ \label{2:Rqts}
\wit{R}(\vec{q},t,s) = \left.
\frac{\delta\langle\wit{S}(\vec{q},t)\rangle}{\delta \wit{h}(\vec{q},s)}
\right|_{h_{\vec{r}}=0}
= e^{-\omega(\vec{q})(t-s)} \sqrt{\frac{g(s)}{g(t)}}
\EEQ
{}From these expressions, the autocorrelation function
$C(t,s) = C_{\vec{0}}(t,s)$ and the autoresponse function
$R(t,s)=R_{\vec{0}}(t,s)$
can be obtained by integrating over the momentum $\vec{q}$.

Summarising, the physically interesting correlation and response functions
are given by equations (\ref{eq12},\ref{eqC},\ref{2:Rqts}) together with
the constraint eq.~(\ref{eq15}), for any time-dependent temperature $T=T(t)$.
This constitutes the main result of the general formalism. 

In order to solve eq.~(\ref{eq15}) explicitly, we now use the specific
form eq.~(\ref{eq4})
of the time-dependent temperature $T=T(t)$. Through a Laplace
transformation
\BEQ \label{2:fA}
\lap{f}(p) = \int_{0}^{\infty} \!\D t\, f(t) e^{-pt}
\EEQ
eq.~(\ref{eq15}) is transformed into a linear difference equation
\BEQ
\lap{g}(p) = \lap{f}(p) + 2T_1\lap{f}(p)\lap{g}(p) + 2(T_0 - T_1)
\lap{f}(p)\lap{g}(p+1/\tau)
\label{eq17}
\EEQ
The analytic solution of this equation allows to study the behaviour of the
correlation and response functions defined above.
Two limit cases correspond to constant temperatures and have been analysed
in detail in the literature \cite{Godr00b}.
First, in the limit $\tau\to\infty$, we have a constant temperature $T=T_0$ and
the final temperature $T_1$ is never reached. Second, the limit
$\tau\to 0$ corresponds to an infinitely rapid quench to the constant end
temperature $T=T_1$. We are interested in the behaviour between
these two extremes.
The solution $\lap{g}(p)$ of eq.~(\ref{eq17}) can be cast in the form 
\BEQ
\lap{g}(p) = \sum_{n=0}^{\infty} {(T_0 - T_1)^n\, \overline g_n (p)}
\label{eq18}
\EEQ
where
\BEQ
\overline g_n (p) =  2^n \prod_{k=0}^n \lap{g}^{(0)}
\left(T_1, ~p+ \frac{k}{\tau}\right)
\label{eq19}
\EEQ
Here $\lap{g}^{(0)}\left(T_1,p\right)$ is the solution as given by
Godr\`eche and Luck \cite{Godr00b} for a fixed temperature $T=T_1$
\BEQ \label{eq20}
\lap{g}^{(0)}(T_1,p) =  \frac{\lap{f}(p)}{1-2T_1\lap{f}(p)}
\EEQ
We shall turn in the next section to a detailed analysis of the properties
of the physical observables coming from this solution.

%%%%%%%%%%%%%%%%%%%%%%%%%%%%%%%%%%%%%%%%%%%%%%%%%%%%%%%%%%%%%%%%%%%%%%%%%%%%%%%%
\section{The equal-time correlator}
%%%%%%%%%%%%%%%%%%%%%%%%%%%%%%%%%%%%%%%%%%%%%%%%%%%%%%%%%%%%%%%%%%%%%%%%%%%%%%%%

We now analyse the long-time behaviour of the equal-time spin-spin correlation
function $C_{\vec{x}}(t)$ or rather its Fourier transform $\wit{C}(\vec{q},t)$.
Throughout this paper, we shall restrict ourselves to the situation where
{\em both} the initial temperature $T_0$ and the final temperature $T_1$ are
above the critical point and the systems is cooled from $T_0$ to $T_1$, viz.
\BEQ
T_0 > T_1 > T_c
\EEQ
The equilibrium critical temperature is given by
$T_c = (2 \lap{f}(0))^{-1}$, see \cite{Coni94,Cugl95,Godr00b}, and is
non-zero for $d>2$.
As we have seen in the last section, the time-dependence of the correlators
follows from the form of the function $g(t)$. In turn, the behaviour of
$g(t)$ can be described in terms of the singularities of its Laplace transform
$\lap{g}(p)$. From eqs.~(\ref{eq18},\ref{eq19}), these singularities are
entirely given in terms of the singularities of the constant-temperature
solution $\lap{g}^{(0)}(T_1,p)$ of eq.~(\ref{eq20}).

In the high-temperature phase, the singularities of $g^{(0)}(T_1,t)$
can be analysed as follows \cite{Godr00b}.
Consider the function
\BEQ
\lap{f}(p) =  \frac{1}{(2\pi)^d}\int_{{\cal B}}\!\D \vec{q}\:
\frac{1}{p + 2\omega(\vec{q})}
\label{eq21}
\EEQ
which is monotonously decreasing with $p$. Furthermore, $\lap{f}(p)$
is analytic in the complex $p$-plane except for a cut in the interval
$-8d\leq p \leq 0$. On the other hand, if
$T_1 \ge T_c$, the function $\lap{g}^{(0)}(T_1,p)$ has a simple pole at a
value $p_0$ given by $\lap{f}(p_0)= 1/(2T_1)$. This translates into an
exponential long-time behaviour of $g^{(0)}(T_1,t)\sim \exp(t/\tau_{\rm eq})$
which defines an equilibration time
\BEQ
\tau_{\rm eq} := \tau_{\rm eq}(T_1) = \tau_0 = \frac{1}{p_0}
\label{eq22}
\EEQ

Now, for a time-dependent temperature $T=T(t)$, the only possible singularities
of $\lap{g}(p)$ are those of $\lap{g}^{(0)}(T_1,p+k/\tau)$ and therefore
the singularities of $\lap{g}(p)$ with a positive real part are simple
poles and occur at
\BEQ
p_k :=p_0- \frac{k}{\tau}
\label{eq23}
\EEQ
with $k=0,1,2,\ldots,k_{\rm max}$. Here $k_{\rm max}$ is the
largest integer such that $p_{k_{\rm max}}$ is still positive.
For $k>k_{\rm max}$, the singularities
merely lead to exponentially decreasing corrections in $g(t)$ and do not 
contribute to the leading long-time behaviour we are interested in.
The sequence of poles $p_k$ leads to the following sequence of relaxation
times
\BEQ
\frac{1}{\tau_{k}(T_1)} = \frac{1}{\tau_{\rm eq}(T_1)}- \frac{k}{\tau}
\label{eq24}
\EEQ 
The quantities $p_k$ are positive for $0\leq k\leq k_{\rm max}$
and negative for $k> k_{\rm max}$.
Hence for $t$ sufficiently large, i.e. $t\gtrsim\tau_{\rm eq}(T_1)$ the leading
contributions to $g(t)$ are of the form
\BEQ
g(t)\simeq  \sum_{k=0}^ {k_{\rm max}} G_k ~\exp(t/\tau_k)
\label{eq25}
\EEQ
where
\BEQ \label{eq26}
G_k = \gamma \sum_{n=0}^{\infty} 2^n \left( T_0 -T_1\right)^n
\prod_{j=0,j\ne k}^{n} \lap{g}^{(0)}\left(T_1, p_0 + (j-k)/\tau\right)
\EEQ
and in addition
\BEQ
\gamma = - \left. \lap{f}(p_0)^2\right/ {\lap{f}\,}'(p_0)
\EEQ
is a positive constant and the prime denotes the derivative.

Using the expression eq.~(\ref{eq25})
the spin-spin correlation function given by eq.~(\ref{eq12})
takes the following form
\BEA
\lefteqn{
\wit{C}(\vec{q},t) \simeq 
{T_1}\left(\sum_{k=0}^{k_{\rm max}}G_k \exp(-kt/\tau)\right)^{-1}
}
\nonumber \\
&\times& \left[ \frac{G_0}{\omega(\vec{q})+\lambda_{\rm eq}^{-2}}
+ \sum_{\ell=1}^{k_{\rm max}} \frac{H_{\ell}}{\omega(\vec{q})
+\lambda_{\ell}^{-2}}
\, e^{-\ell t/\tau} + \frac{T_0-T_1}{T_1}
\frac{G_{k_{\rm max}}}{\omega(\vec{q})-\lambda_{k_{\rm max}+1}^{-2}}
\, e^{-(k_{\rm max}+1)t/\tau} \right]
\label{eq27} \\
&+& O\left( e^{-2\omega(\vec{q})t}\right)
\nonumber
\EEA
provided $t\gtrsim\tau_{\rm eq}(T_1)$ and
where for all $0\leq \ell\leq k_{\rm max}+1$ we have defined
\BEQ \label{eqlambda}
\lambda_{\ell} = +\sqrt{2|\tau_{\ell}|\,}=
\left| \frac{2\tau_{\rm eq}(T_1)\tau}
{\tau-\ell \tau_{\rm eq}(T_1)}\right|^{1/2}
\EEQ
In the following, we shall use the relation
$\lambda_{\rm eq}=\lambda_0=\sqrt{2\tau_{\rm eq}}$. In addition, we also used
the abbreviation
\BEQ
H_k = G_k + \frac{T_0-T_1}{T_1} G_{k-1}
\EEQ

In order to understand the behaviour of $\wit{C}(\vec{q},t)$, we note that
the first term in eq.~(\ref{eq27}), which formally corresponds to $\ell=0$,
gives the stationary contribution. In addition,
there is at least one additional transient term which will disappear in the
limit $t\gg \tau$.
The number of transient terms which are present depends
on the value of $k_{\rm max}=k_{\rm max}(T_1,\tau)$. We now discuss the various
physical cases which can arise.

\subsection{First-mode contribution -- Short thermal relaxation times}
\label{zero}
\begin{figure}[hb]
\centerline{\epsfxsize=4.75in\epsfbox
{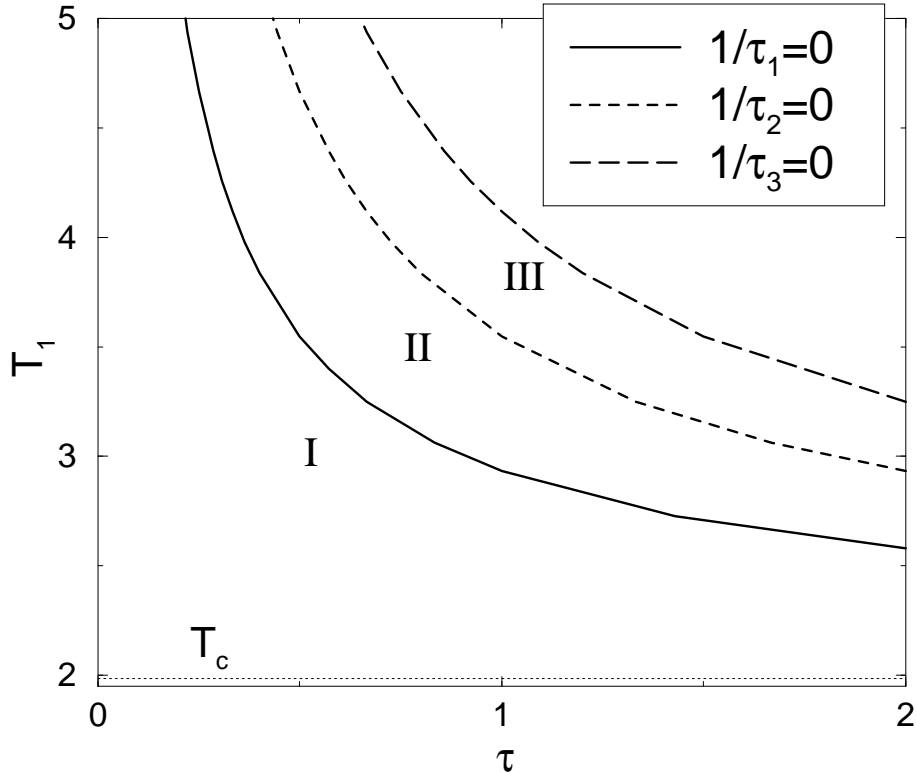}}
\caption{Kinetic phase diagram of the $3D$ spherical model with
a time-dependent temperature as given by {\protect eq.~(\ref{eq4})}.
The lines shown are the loci where the relaxation times $\tau_{\ell}$,
$\ell=1,2,3$ diverge. The dotted line marked $T_c$ gives the equilibrium
critical point and the regions I, II and III are those with
$k_{\rm max}=0,1,2$, respectively.
\label{Abb1}}
\end{figure}

The simplest case arises if $\tau_{\rm eq}\ge \tau$, that is the bath relaxes
faster than the system itself. Then $p_0 - \frac{1}{\tau}\le0$,
that is $k_{\rm max}=0$ and we have in eq.~(\ref{eq27}) $\ell = 0$ as the
only contributing mode.
Therefore (assuming $t> \tau$ and $t\gg \tau_{\rm eq}(T_1)$)
\BEQ
\wit{C}(\vec{q},t) \simeq T_1\left(\frac{1}{\omega(\vec{q})
+\lambda_{\rm eq}^{-2}(T_1)}
+ \frac{(T_0-T_1)}{T_1} \frac{e^{-t/\tau}}
{\omega(\vec{q})-\lambda_1^{-2}}\right)
+ O\left( e^{-2\omega(\vec{q})t-t/\tau}\right)
\label{eq28}
\EEQ
For a physical understanding, it might be more appealing to consider
the correlator in real space. In the $1D$ case, the result is particularly
simple and becomes in the large-separation limit $|\vec{x}|\gg 1$
\BEQ \label{eqCx}
C_{\vec{x}}(t) \simeq T_1 \left( \lambda_0 e^{-|\vec{x}|/\lambda_0}
-\frac{T_0-T_1}{T_1} \lambda_1
\sin\left(\frac{|\vec{x}|}{\lambda_1}\right)e^{-t/\tau} \right)
+ O\left( e^{-\vec{x}^2/t}e^{-t/\tau}\right)
\EEQ
Therefore,
the correlator is the sum of a stationary and spatially homogeneous
term and a transient term which in addition shows spatial oscillations
of wavelength $\lambda_1$. We point out that $\lambda_1$
diverges when the external time scale $\tau$ approaches the internal
equilibration time $\tau_{\rm eq}$ from below, see eq.~(\ref{eqlambda}).
In figure~\ref{Abb1}
we show the location of the line $\tau_{\rm eq}=\tau$ as the curve $1/\tau_1=0$
in the $(T_1,\tau)$-plane. The region I in figure~\ref{Abb1}
corresponds to the case $k_{\rm max}=0$.

For larger values of $\tau$ we go over to a multimode regime which we
discuss below.

%%----------------------------------------------------------------------------%%
%%----------------------------------------------------------------------------%%

\subsection{Multimode contributions}

If $k_{\rm max} > 0$ several transient modes contribute to
$\wit{C}(\vec{q},t)$. Their contributions are quite different, however.
The propagator arising in all terms is of the form
\BEQ
\frac{1}{\omega(\vec{q}) \pm \lambda_k^{-2}}
\stackrel{q\to 0}{\simeq}\frac{1}{q^{2} \pm \lambda_k^{-2}}
\label{eq29} 
\EEQ 
with a positive sign for $k\leq k_{\rm max}$ and a negative sign
for $k=k_{\rm max}+1$. This leads to a different spatial
behaviour, in analogy with eq.~(\ref{eqCx}) found for the case
$k_{\rm max}=0$ discussed above. The first $k_{\max}$ terms all lead to
spatially decaying contributions to $C_{\vec{x}}(t)$ in the large-$|\vec{x}|$
limit, where the characteristic length scale is given by $\lambda_{\ell}$
with $\ell=0,1,\ldots,k_{\rm max}$. With the exception of the $\ell=0$
contribution, all terms are transient and decay with a relaxation time given
by $\tau/{\ell}$. In contrast, the last term, which is
also transient, gives rises to a spatially oscillating term with a wave length
$\lambda_{\rm osc} := \lambda_{k_{\rm max}+1}$.

The behaviour of the system in the first few regimes is illustrated in
figure~\ref{Abb1}. For a fixed final temperature $T_1$ and increasing values of
the external relaxation time $\tau$, the system goes over from region I
with only a single mode (since $k_{\rm max}=0$) to region II with modes,
later to region III with three modes and so on.

%%----------------------------------------------------------------------------%%
\begin{figure}[hb]
\centerline{\epsfxsize=4.75in\epsfbox
{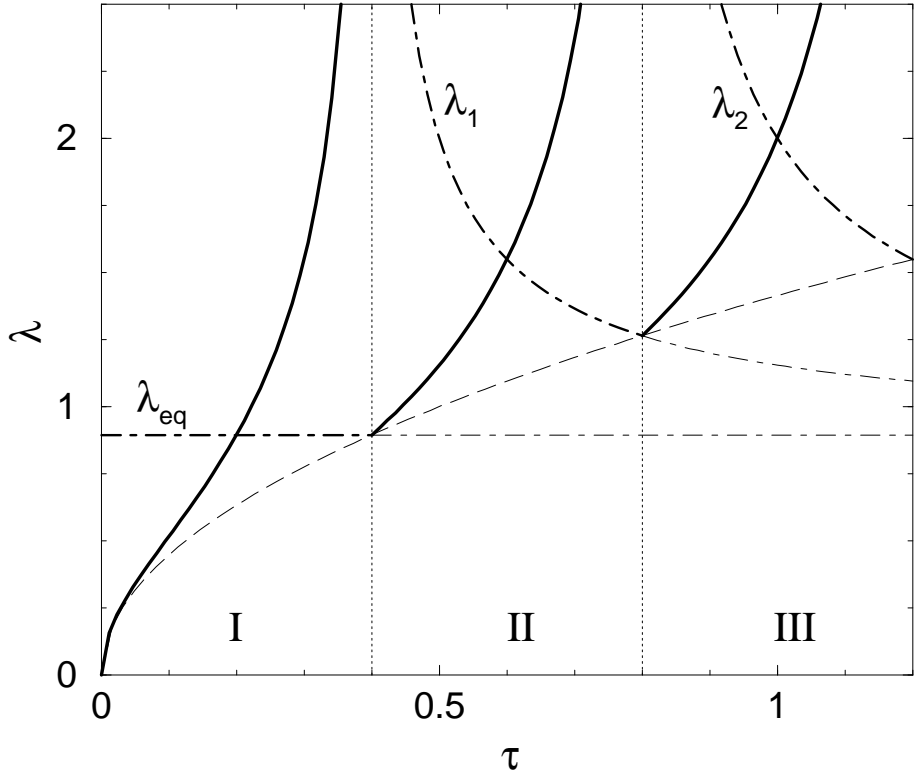}}
\caption{Dependence of some correlation and wave lengths $\lambda_{\ell}$
on the external relaxation time $\tau$, at fixed final temperature
$T_1=3.8363\ldots$ in the
$3D$ spherical model. The regions I, II, III correspond to the cases
$k_{\rm max}=0,1,2$, respectively. The thick solid lines give the oscillation
wave length $\lambda_{\rm osc}=\lambda_{k_{\rm max}+1}$. The dash-dotted
lines give the equilibration length scale $\lambda_{\rm eq}=\lambda_0$ 
(the equilibration time $\tau_{\rm eq}\sim \lambda_{\rm eq}^2$) and the
first additional length scales $\lambda_1$ and $\lambda_2$. The dashed line
corresponds to the curve $\lambda=\sqrt{2\tau}$, see text.
\label{Abb2}}
\end{figure}
%%----------------------------------------------------------------------------%%

The behaviour of the various correlation and wave lengths is illustrated in
figure~\ref{Abb2} for the $3D$ case and for $\tau_{\rm eq}=0.4$, that is
$T_1=3.8363\ldots$. First, we note that the characteristic length scale
$\lambda_{\rm eq}$ associated with the only non-transient term does {\em not}
depend on the externally imposed time scale $\tau$. Second, the thick solid
lines show the dependence of the oscillation wave length
$\lambda_{\rm osc}=\lambda_{k_{\rm max}+1}$ on $\tau$. If we
start with a small value of $\tau$ in region I and then increase $\tau$,
$\lambda_{\rm osc}$ increases and finally diverges when $\tau=\tau_{\rm eq}$
that is when the external time scale becomes equal to the internal equilibration
time scale. If $\tau$ is now increased further, we enter into region II.
At this point, the previous oscillating mode transforms into a spatially
monotonous one which simply gives another contribution with decreases as a
function on $|\vec{x}|$ and in addition,
a new oscillating mode with a {\em finite}
wave length appears. This new mode arise since one of the 
modes which in region I
leads to exponentially decaying contribution to $g(t)$ transforms itself into
an, albeit non-leading, exponentially growing contribution. Furthermore, 
the characteristic length of the old oscillating mode (denoted by $\lambda_1$
in figure~\ref{Abb2}) now decreases as a function of $\tau$. We now increase
$\tau$ further until the oscillation wave length $\lambda_{\rm osc}$ diverges
again and we therefore go over from region II into region III. At this
transition point, we observe that the value of $\lambda_1$ agrees with the one
of the new oscillation wave length for region III. Phenomenologically, this
looks as if that mode would split into two modes, one giving rise to a
spatially oscillating contribution to $C_{\vec{x}}(t)$ and the other one to
a spatially decaying term. This splitting occurs along the line
$\lambda=\sqrt{2\tau}$ and in particular, it follows that the oscillation
length always satisfies $\lambda_{\rm osc}\geq \sqrt{2\tau}$.

%%%%%%%%%%%%%%%%%%%%%%%%%%%%%%%%%%%%%%%%%%%%%%%%%%%%%%%%%%%%%%%%%%%%%%%%%%%%%%%%
\section{Two-time correlations and response functions}
%%%%%%%%%%%%%%%%%%%%%%%%%%%%%%%%%%%%%%%%%%%%%%%%%%%%%%%%%%%%%%%%%%%%%%%%%%%%%%%%

Using the expression of the function $g(t)$ from eq.~(\ref{eq25}),
we can formally compute the two-time observables we are interested in, that is
the two-time autocorrelation function and the autoresponse function defined in
eqs.~(\ref{eqC},\ref{2:Rqts}). In particular, a study of these two
functions allows to calculate the fluctuation-dissipation ratio $X(t,s)$.
A deviation of that ratio from unity should signal a departure from
thermodynamic equilibrium. For a sudden quench to a final temperature $T_1$
this question has already been studied \cite{Godr00b} and it was shown that
indeed $X(t,s)=1$ in the long-time limit in the high-temperature phase
when $T_1 > T_c$. Here, with the additional complication of a time-dependent
heat bath temperature, we must first define an instantaneous time-dependent
temperature of the system in order to make the criterion of the
fluctuation-dissipation ratio applicable. We choose the equilibrium part 
$\wit{C}_{\rm eq}$ of the single-time correlator as the physical
quantity which measures a temperature. Consider for a moment the case of a
fixed temperature $T$. The equilibrium part behaves for small
momenta (or large distances) as
\BEQ
\lim_{\vec{q}\to 0} \wit{C}_{\rm eq}(\vec{q}) = \lim_{\vec{q}\to 0}
\frac{T}{\omega(\vec{q})+\lambda_{\rm eq}^{-2}} = 2 \tau_{\rm eq} T
\EEQ
and depends on $T$ in a known way. Therefore, we use the extension of this
relation to {\em define}, for a time-dependent bath temperature
$T(t)$ varying according to eq.~(\ref{eq4}), an effective
temperature $T_{\rm eff}(s)$ of the spin system.
Taking the explicit form (\ref{eq27}) of the single-time propagator
$\wit{C}(\vec {q},s)$ we set
\BEQ
\lim_{\vec{q}\to 0} \wit{C}(\vec {q},s) = 2\tau_{\rm eq}(T_1)T_{\rm eff}(s)
\EEQ
which is the defining equation for the effective time-dependent 
temperature $T_{\rm eff}(s)$. It remains to be seen whether this $T_{\rm eff}$
can play the role of a physical equilibrium temperature.

This can be seen by means of a comparison with the results
obtained from an instantaneous quench towards $T_1$, hence $\tau=0$.
For clarity, we shall denote by $A^{(0)}$ the result obtained for an observable
$A$, after a sudden quench with $\tau=0$ to the final temperature $T_1$.
The two-time autocorrelation function $C^{(0)}(t,s)$ is explicitly given
in eq. (2.58) of \cite{Godr00b}.
In particular, it is known that \cite{Godr00b}
$\lim_{\vec{q}\to 0}\wit{C}^{(0)}(\vec {q},s) =2\tau_{\rm eq}(T_1)T_{1}$.

We now consider the two-time autocorrelation function $C(t,s)$.
From section~2, we have
\BEA
C(t,s) &=& \sqrt{\frac{g(s)}{g(t)}} \int \!\D\vec{q}\: \wit{C}(\vec{q},s)
e^{-\omega(\vec{q})(t-s)}
\nonumber \\
&\simeq&  \sqrt{\frac{g(s)}{g(t)}} \int \!\D\vec{q}\: \wit{C}(\vec{0},s)
e^{-\vec{q}^2(t-s)}
\nonumber \\
&=& \sqrt{\frac{g(s)}{g(t)}} \int \!\D\vec{q}\: 2 \tau_{\rm eq}(T_1)
T_{\rm eff}(s) e^{-\vec{q}^2(t-s)}
\nonumber \\
&=&  \sqrt{\frac{g(s)}{g(t)}} \int \!\D\vec{q}\: \frac{T_{\rm eff}(s)}{T_1}
\wit{C}^{(0)}(\vec{0},s) e^{-\vec{q}^2(t-s)}
\nonumber \\
&\simeq& \frac{T_{\rm eff}(s)}{T_1} k^{(0)}(t,s) C^{(0)}(t,s)
\label{eq:Ctsgrand}
\EEA

In the second and the fifth line we used the fact that we are in the regime
$t-s\gg 1$, in combination with the fact that in this limit, only
the low-$\vec{q}$ behaviour of $C(t,s)$ and of $C^{(0)}(t,s)$ really
contributes. In the fourth line, the definition of $T_{\rm eff}(s)$ was
used. For brevity, we have set
\BEQ
k^{(0)}(t,s)= \sqrt{\frac{g(s)/g^{(0)}(s)}{g(t)/g^{(0)}(t)}}
\EEQ
where the Laplace transform of $g^{(0)}(t)$ was given in (\ref{eq20}).
The exact autoresponse function is given by
\BEQ
R(t,s)=\int \!\D \vec{q}\: \wit{R}(\vec{q},t,s) =
f\left(\frac{t-s}{2}\right)\sqrt{\frac{g(s)}{g(t)}} =
R^{(0)}(t,s) k^{(0)}(t,s)
\label{eqRtau}
\EEQ

We now ask whether the quasi-equilibrium temperature $T_{\rm eff}(s)$ so 
defined is really the temperature of a system in equilibrium. A convenient way
to test this is through the fluctuation-dissipation ratio
\BEQ
X(t,s)=T_{\rm eff}(s)R(t,s)\left(\frac {\partial C(t,s)}{\partial s}\right)^{-1}
\label{eqX}
\EEQ
An equilibrium situation can only be recovered if $X(t,s)=1$.
We now bring this expression into a simpler form as follows
\BEA
\frac{1}{X(t,s)} &=& \frac{\partial C(t,s)/\partial s}{T_{\rm eff}(s) R(t,s)}
\nonumber \\
&=& \frac{1}{T_{\rm eff}(s)} \frac{C(t,s)}{R(t,s)}
\frac{\partial C(t,s)/\partial s}{C(t,s)}
\nonumber \\
&\simeq& \frac{1}{T_{\rm eff}(s)}
\frac{C^{(0)}(t,s) T_{\rm eff}(s) T_1^{-1}
k^{(0)}(t,s)}{R^{(0)}(t,s) k^{(0)}(t,s)}
\frac{\partial}{\partial s} \ln C(t,s)
\nonumber \\
&=& \frac{1}{T_1} \frac{C^{(0)}(t,s)}{R^{(0)}(t,s)}
\frac{\partial}{\partial s} \ln C(t,s)
\nonumber \\
&\simeq& \left( \frac{\partial}{\partial s} \ln C(t,s)\right)\,
\left(\frac{\partial}{\partial s} \ln C^{(0)}(t,s) \right)^{-1}
\EEA
In performing this calculation, we used in the third line the approximate
expressions eq.~(\ref{eq:Ctsgrand},\ref{eqRtau})
valid for $t-s\gg \tau_{\rm eq}(T_1)$
and in the last line, we used the fact the
fluctuation-dissipation theorem holds
for any instantaneous quench to the final temperature $T_1>T_c$ \cite{Godr00b}.

After some algebra, we can write this in a compact form,  
taking also the $t\to\infty$ limit
\BEQ
X(t,s)=\left[1+\frac{\tau_{\rm eq}}{\tau}
\Phi\left(\frac{s}{\tau}\right)\right]^{-1}
\label{eqX2}
\EEQ
Remarkably, this scaling form only depends on the single scaling variable
$x=s/\tau > 1$.
The associated scaling function reads
\BEQ
\Phi(x) = \left( \frac{\partial}{\partial x} \ln \Phi_1(x)
+ 2 \frac{\partial}{\partial x} \ln \Phi_2(x) \right)
\EEQ
where explicitly
\BEA
\Phi_1(x) &=& \sum_{k=0}^{k_{\rm max}} \frac{G_{k}}{G_0}\, e^{-k x}
\\
\Phi_2(x) &=& T_1 \left[ \sum_{\ell=0}^{k_{\rm max}}
G_{\ell}\, e^{-\ell x} \right]^{-1}
\nonumber \\
& & \times \left[ G_0 \lambda_{\rm eq}^2 + \frac{T_0-T_1}{T_1} G_{k_{\rm max}}
\lambda_{k_{\rm max}+1}^2 e^{-(k_{\rm max}+1)x} +
\sum_{k=1}^{k_{\rm max}} H_k \lambda_k^2 e^{-k x}\right]
\EEA

A particularly simple form is found for $k_{\rm max}=0$. Physically, this
corresponds to the situation when $\tau\leq\tau_{\rm eq}(T_1)$.
Then the fluctuation dissipation ratio reads
\BEQ
X(t,s)=\left[1+2\frac{\tau_{\rm eq}}{\tau}\left(\frac{\lambda_{0}^{2}T_1}
{\lambda_{1}^{2}(T_0-T_1)}\exp(x)-1\right)^{-1}\right]^{-1}
\;\; , \;\; x = \frac{s}{\tau}
\label{eqX3}
\EEQ
As in the case of an infinitely rapid quench, the fluctuation-dissipation
ratio is unity, up to corrections which vanish exponentially fast in the
long-time limit. We can conclude that the system evolves towards
thermodynamic equilibrium.

%%%%%%%%%%%%%%%%%%%%%%%%%%%%%%%%%%%%%%%%%%%%%%%%%%%%%%%%%%%%%%%%%%%%%%%%%%%%%%%%
\section{The clamped case and the limit $\tau/\tau_{\rm eq}(T_0)
\rightarrow \infty$}
%%%%%%%%%%%%%%%%%%%%%%%%%%%%%%%%%%%%%%%%%%%%%%%%%%%%%%%%%%%%%%%%%%%%%%%%%%%%%%%%

Studying the kinetics with a time-dependent temperature and in the disordered
phase, way may identify the following time regimes:
\begin{enumerate}
\item $\tau\ll \tau_{\rm eq}(T_1)$. This case formally corresponds
to $k_{\rm max}=0$. Studying the long-time kinetics, we always work in the
situation where $t\gg\tau_{\rm eq}(T_1)\gg\tau$. Up to exponentially small
corrections, we recover an equilibrium dynamics as found
for an infinitely rapid quench \cite{Godr00b}.
\item $\tau \simeq \tau_{\rm eq}(T_1)$. In this case, we consider
$t\gg\tau_{\rm eq}(T_0)$ and also $t\gg\tau$. This regime corresponds
to a finite non-zero value of $k_{\rm max}$. With respect to the situation
of an infinitely rapid quench, there are several new exponentially small 
contributions which arise, as was shown in section 4. 
As expected, equilibrium dynamics is recovered.
\item $\tau\gg \tau_{\rm eq}(T_1)$. 
This is called the clamped regime\footnote{en fran\c{c}ais: serr\'e;
auf deutsch: eingeklemmt}  because the
microscopic degrees of freedom follow the varying temperature of the external
heat bath. Certainly, for times 
$t\gg \tau\gg \tau_{\rm eq}(T_0)$ we shall simply recover equilibrium dynamics,
along the lines of section 4. Because we consider only quenches which
reduce the temperature $T_0 > T_1 > T_c$ throughout, we always have
$\tau_{\rm eq}(T_1)>\tau_{\rm eq}(T_0)$. On the other hand, 
new and interesting long-time behaviour
occurs in the regime $\tau_{\rm eq}(T_0)\ll t \ll \tau$. In this case, 
the system is in equilibrium as far as the internal degrees of freedom are
concerned, since $t\gg \tau_{\rm eq}(T_0)$, but has not yet equilibrated with 
respect to the external time scale of the heat bath, since $t\ll \tau$.
(a special case of the case under consideration here corresponds to 
$k_{\rm max}=\infty$.)
\end{enumerate}

We first have to look closer at the limit $\tau\to\infty$.
As we have seen above,
the central part of the calculation is the solution of the Volterra
integral equation (\ref{eq15}) coming from the spherical constraint. However,
for $\tau\to\infty$ and after having performed the Laplace transformation,
we may expand in $1/\tau$ and then obtain to leading order an ordinary
differential equation
\BEQ \label{eqdgl}
\frac{\D \lap{g}(p)}{\D p} + \tau a(p) \lap{g}(p) = \tau b(p)
\EEQ
where
\BEQ \label{eqab}
a(p) = \frac{2T_0 \lap{f}(p) -1}{2(T_0-T_1)\lap{f}(p)} \;\; , \;\;
b(p) = -\frac{1}{2} \frac{1}{T_0-T_1}
\EEQ
Although this equation may be solved straightforwardly through
integrations, it is for our purposes more useful to proceed differently.
After all, we are merely interested in the large-time (or small-$p$) behaviour
of the correlators. In addition, because of the $\tau\to\infty$ limit
implicit in eq.~(\ref{eqdgl}), we must take the limit in such a way that 
$\tau a(p)$ remains finite. Therefore, we merely need $a(p)$ in the
vicinity of
$p_c$ defined by $a(p_c)=0$.\footnote{Physically, one may understand this
by noting that in the case of constant temperature $T_0$, $\tau$
is formally infinite and we should have to leading order
$\lap{g}(T_0,p)=b(p)/a(p)$. Therefore, the root $p_c$ sets the entire
physical behaviour, up to normalization factors.}
The leading long-time behaviour should then be
given by an expansion in powers of $p-p_c$. We write, using (\ref{eqab})
\BEQ
a(p) \simeq \Delta(p-p_c) + \ldots \;\; , \;\;
\Delta = \frac{2 T_0^2}{T_0-T_1}
\left.\frac{\D \lap{f}(p)}{\D p}\right|_{p=p_c} < 0
\EEQ
We look for a solution of (\ref{eqdgl}) of the form
\BEQ
\lap{g}(p)=\sum_{n=0}^{\infty} {\tau}^{-n}\lap{g}_{n}(p)
\label{dev}
\EEQ
and we directly obtain
\BEQ \label{gn}
\lap {g}_{n}(p)=\frac{b}{\Delta}\left(\frac{2}{\Delta}\right)^n
\frac{\Gamma(n+1/2)}{\Gamma(1/2)}
(p-p_c)^{-2n-1}
\EEQ
In order to perform the inverse Laplace transformation, we write
\BEQ
\lap{g}(p) = \sum_{n=0}^{\infty} \tau^{-n}\,
\int_{0}^{\infty} \!\D t\, g_n(t) e^{-pt}
= \sum_{n=0}^{\infty} \int_{0}^{\infty} \!\D t\, \frac{b}{\Delta}
\left(\frac{t^2}{2\tau\Delta}\right)^n \frac{1}{n!}\: e^{-(p-p_c)t}
\EEQ
At this stage the sum may be performed first and we find
\BEQ
g(t)=\sum_{n=0}^{\infty} g_n(t) e^{p_c t} =
\frac{b}{\Delta}\exp\left(p_c t -\frac{t^2}{\tau_{\rm eff}^2}\right)
\EEQ
which is valid provided $t\ll \tau$.
In this expression, $\tau_{\rm eff}$ is the typical
time after which the function $g$
and then the correlators which are closely related to it differ notably from
the expressions they take at formally infinite external relaxation time $\tau$.
We find
\BEQ
\tau_{\rm eff}=\sqrt{-2\tau\Delta}=\sqrt{2\tau \frac{2 T_0^2}{T_0-T_1}
\left|\frac{\D \lap{f}(p)}{\D p}\right|_{p=p_c}}
\label{eqtaueff}
\EEQ
Remarkably enough, $\tau_{\rm eff}$ differs from the external time
$\tau_{\rm ext}=\tau$. Rather, it is the geometric mean
between the externally imposed relaxation time scale as measured by $\tau$
and an internal time scale, measured by $\tau_{\rm dep}=2\Delta(T_0,T_1)$.
In the physical observables, $\tau_{\rm eff}$ appears as a natural
time scale informing on the departure from an effectively constant temperature
$T=T_0$. We point out that this new time scale $\tau_{\rm dep}$ is
distinct from the equilibration time at temperature $T_0$, namely
$\tau_{\rm eq}(T_0)=1/p_c$.

For example, the two-time response function becomes
\BEQ
R(t,s) = R^{(0)}(t,s) \exp\left( - \frac{(t-s)(t+s)}{4\tau |\Delta|}\right)
\EEQ
which illustrates again the role of $\tau_{\rm eff}$ in the description
of the deviation with respect to $R^{(0)}(t,s)$.

We now comment on how these results depend on spatial dimensionality $d$.
{}From eq.~(\ref{eqtaueff}), we directly see that $\tau_{\rm eff}$ diverges
when $T_0$ and $T_1$ get closer to each other. Far from the critical region,
we have $\tau_{\rm eff}\simeq(T_0-T_1)^{-1/2}$, independently of $d$.
On the other hand, when the system enters the critical region, results depend
on whether fluctuations are strong as is the case for $2<d<4$ or whether
one has mean-field criticality when $d>4$. In
the fluctuation-dominated regime, that is $2<d<4$,
the first derivative of $\lap{f}(p)$ diverges as $p$ goes to zero.
Using the known small-$p$ expansion of $\lap{f}$ \cite{Godr00b,Mont65,Luck85},
we find
\BEQ
\tau_{\rm eff}\simeq\left[\frac{T_1-T_c}{T_c}\right]
^{-\frac{4-d}{2(d-2)}}(T_0-T_1)^{-1/2}
\label{taueff}
\EEQ
On the other hand, in the mean-field regime $d>4$, the first derivative of
$\lap{f}(p)$ remains bounded even in the neighbourhood of the critical point.
Then we simply retain $\tau_{\rm eff}\simeq(T_0-T_1)^{-1/2}$.

For $2<d<4$, we have thus found a rather counterintuitive effect. Namely, the
time scale $\tau_{\rm eff}$ up to which the system evolves as if the temperature
were fixed at $T_0$, increases when the final temperature $T_1$ approaches the
critical temperature $T_c$. This enhancement of $\tau_{\rm eff}$ is
absent for $d>4$ and must therefore be related to the presence of strong
fluctuation effects close to $T_c$ and below the upper critical dimension.

Finally, we consider the two-time observables and the
fluctuation-dissipation ratio. We have just seen that for $\tau$ large,
the kinetics of the system occurs at an infinitesimally slow
rate for a long time. In contrast to the treatment of section 4, it is
therefore useful to compare the time-evolving
quantities with the ones when formally $\tau=\infty$ or, in physical terms,
when the temperature remains fixed at $T_0$. Formally, the calculations
can be done by analogy with section 4.
First, we use the single-time correlator to define an effective time-evolving
temperature $T_{\rm eff}(s)$ from the relationship
\BEQ \label{5:Teff}
\lim_{\vec{q}\to 0} \wit{C}(\vec {q},s) = 2\tau_{\rm eq}(T_0)T_{\rm eff}(s)
\EEQ
We shall denote by $A^{(\infty)}$ the result obtained for an observable $A$
with a time evolution governed by an infinitely long external time
$\tau=\infty$. In particular, we have
$\lim_{\vec{q}\to 0} \wit{C}^{(\infty)}(\vec{q},s) = 2 T_0\tau_{\rm eq}(T_0)$.
As in section 4, it is easy to see that
\BEQ
C(t,s) = \frac{T_{\rm eff}(s)}{T_0} k^{(\infty)}(t,s) C^{(\infty)}(t,s)
\;\; , \;\;
k^{(\infty)}(t,s) = \sqrt{ \frac{g(s)/g^{(\infty)}(s)}{g(t)/g^{(\infty)}(t)} }
\EEQ
where $g^{(\infty)}(t)$ is again the solution of the spherical constraint
(\ref{eq15}) but now for $\tau=\infty$ and therefore $T_1$ is replaced by
$T_0$. 

We now want to find $T_{\rm eff}(s)$ explicitly.
Using the general expressions for the
single-time correlator $\wit{C}(\vec{q},t)$ from section 2, the explicit form
of $g(t)$ found above and taking the limit $\vec{q}\to 0$, we obtain
\BEQ
T_{\rm eff}(s) = \frac{\Delta\exp(-p_c s + s^2/\tau_{\rm eff}^2)}
{2\tau_{\rm eq}(T_0)b}
\left[ 1 + \frac{2b}{\Delta} \int_{0}^{s}\!\D t_1\:
\exp\left( p_c t_1 - \frac{t_1^2}{\tau_{\rm eff}^2}\right)
\left( T_1 + (T_0-T_1)e^{-t_1/\tau} \right) \right]
\EEQ
Now, we consider times $s$ which satisfy
\BEQ \label{eq:regime}
\frac{s}{|\Delta|} \ll \frac{\tau}{\tau_{\rm eq}(T_0)}
\EEQ
From the explicit form of $\Delta$, it is clear that this condition is the
easier to satisfy the closer $T_0$ to $T_1$, even if
both are far from the critical point $T_c$.
We then obtain
\BEA
T_{\rm eff}(s) &=& \frac{\Delta e^{-p_c s}}{2b\tau_{\rm eq}(T_0)}
\left[ 1 + \frac{2b}{\Delta} T_1  p_c^{-1} e^{p_c s} +
\frac{2b}{\Delta} (T_0-T_1) \frac{e^{(p_c-1/\tau)s}}{p_c-1/\tau}
+ O\left( e^{-p_c s}\right) \right]
\nonumber \\
&=& T(s)
\EEA
where in the last step we used that $\tau\gg\tau_{\rm eq}(T_0)$ or equivalently
that $p_c\gg 1/\tau$, and we recall the definition $\lap{f}(p_c)=1/T_0$. 
Therefore, we see that under the stated conditions,
the effective temperature is equal to the temperature of the external
heat bath. This nicely confirms that our definition of $T_{\rm eff}(s)$ is a
natural one.

Physically, this result illustrates the (expected) applicability of the 
adiabatic approximation to equal-time two-point correlators. 
Namely, we have seen that in
the limit $\tau\gg \tau_{\rm eq}(T_0)$, the effective temperature 
$T_{\rm eff}(s)$ equals the bath temperature $T(s)$ and the equal-time 
two-point correlators are those of a quasi-equilibrium system.

Lastly, we examine to what extent this heuristic expectation concerning
equal-time correlators carries over to two-time quantities. First, in an
analogous manner to section 4, the two-time response function reads   
\BEQ
R(t,s) = f\left(\frac{t-s}{2}\right) \sqrt{\frac{g(s)}{g(t)}}
= R^{(\infty)}(t,s) k^{(\infty)}(t,s)
\EEQ 
Then, we can easily compute the fluctuation-dissipation ratio $X(t,s)$.
Formally, it is again defined by (\ref{eqX}) and we find as before
\BEQ
\frac{1}{X(t,s)} = \left( \frac{\partial}{\partial s} \ln C(t,s) \right)
\left( \frac{\partial}{\partial s} \ln C^{(\infty)}(t,s) \right)^{-1}
\EEQ
and this holds provided $t-s\gg \tau_{\rm eq}(T_0)$.
In the just-studied adiabatic regime, we have
\BEQ
C(t,s) = \frac{T(s)}{T_0} e^{(s^2-t^2)/\tau_{\rm eff}^2}\, C^{(\infty)}(t,s)
\EEQ
and the fluctuation-dissipation ratio becomes
\BEA \label{eq:ratio}
\frac{1}{X(t,s)} &=& \left( \frac{\partial}{\partial s} \ln \frac{T(s)}{T_0}
+\frac{2s}{\tau_{\rm eff}^2}+ \frac{1}{2\tau_{\rm eq}(T_0)} \right)
2\tau_{\rm eq}(T_0)
\nonumber \\
&=& 1 - \frac{2s}{\Delta}\frac{\tau_{\rm eq}(T_0)}{\tau}
- 2 \frac{\tau_{\rm eq}(T_0)}{\tau}
\frac{ (T_0-T_1)e^{-s/\tau}}{T_1 + (T_0-T_1) e^{-s/\tau}}
\nonumber \\
&\simeq& 1 - 2 \frac{\tau_{\rm eq}(T_0)}{\tau} \frac{T_0-T_1}{T_0}
\EEA
where the last relation is valid provided $s\ll \tau$ and where
the condition eq.~(\ref{eq:regime}) was used.

For an equilibrium system, it is well-known that $X(t,s)=1$ for all times 
$s<t$ from the fluctuation-dissipation theorem. Here, we find an intermediate
regime $\tau_{\rm eq}(T_0)\ll s\ll \tau$ such that $X(t,s)$ reaches a plateau
value which is different from unity. Still, the adiabatic approximation remains
valid, up to small corrections in $s/\tau$. However, the fact that for 
$s\ll \tau$ the difference $X(t,s)-1$ is finite (albeit it may numerically be
small) shows that the validity of the fluctuation-dissipation relation is
independent of, and more restrictive than, the adiabatic approximation. 

Of course, for $s\gg \tau\gg \tau_{\rm eq}(T_0)$ the fluctuation-dissipation
theorem will be recovered. In other words, eq.~(\ref{eq:ratio}) shows that the
equilibration $s\gg \tau_{\rm eq}(T_0)$ with respect to the internal degrees of
freedom is already enough to reproduce the flucutation-dissipation theorem to a
good approximation while it becomes exact when equilibration with respect to the
bath, $s\gg \tau$ is also achieved. 
 
%%%%%%%%%%%%%%%%%%%%%%%%%%%%%%%%%%%%%%%%%%%%%%%%%%%%%%%%%%%%%%%%%%%%%%%%%%%%%%%%
\section{Conclusions}
%%%%%%%%%%%%%%%%%%%%%%%%%%%%%%%%%%%%%%%%%%%%%%%%%%%%%%%%%%%%%%%%%%%%%%%%%%%%%%%%

We have studied how the non-equilibrium kinetics of a simple ferromagnet
is modified when the system is not instantaneously quenched to a final
temperature $T_1$ but brought into contact with a heat bath starting at an
initial temperature $T_0$ such that the approach
towards $T_1$ is described by a finite external relaxation time $\tau$,
according to (\ref{1:eqTt}). More general temperature histories $T(t)$ of the
heat bath can be decomposed into sums of exponentials via Laplace
transformations. Then the physically interesting long-time behaviour will
be governed by the largest relaxation time arising in the Laplace spectrum. 
We have studied as an example the exactly solvable spherical model in order
to get physical insight into the role of a finite value of $\tau$.
We have concentrated on the high-temperature phase such that
$T_0 > T_1 > T_c$, generalizing previous studies
\cite{Coni94,Cugl95,Zipp00,Godr00b,Corb02} with $\tau=0$.

Our results can be summarized as follows:
\begin{enumerate}
\item If $\tau$ is smaller than or at most of the same order of
magnitude as the internal relaxation or
equilibration time $\tau_{\rm int}=\tau_{\rm eq}$,
then for times large compared to $\tau,\tau_{\rm eq}$ 
the system approaches exponentially fast the known stationary state, in
agreement with what was found previously for
$\tau=0$ \cite{Coni94,Cugl95,Zipp00,Godr00b,Corb02}.

In addition, there are transient contributions to the spin-spin
correlation function with spatial oscillations on finite length scales
$\lambda$. In particular, if $\tau \leq \tau_{\rm eq}$, we find
\BEQ
\lambda^{-1} \sim \sqrt{ \frac{1}{\tau}- \frac{1}{\tau_{\rm eq}} }
\EEQ
\item In this case, the stationary state is indeed in thermodynamic
equilibrium and is reached exponentially fast for times larger than $\tau$.
We have checked this through explicit calculation of the fluctuation-dissipation
ratio $X(t,s)\to 1$ for late times.
\item A qualitatively different behaviour may be observed in the clamped case,
where $\tau\gg \tau_{\rm eq}$.

A common physical example with clamped degrees of freedom is a glass.
A system in a glass state remains out of equilibrium as attested by the
fact that the fluctuation-dissipation limit
$X_{\infty}=\lim_{s\to\infty}\lim_{t\to\infty} X(t,s)\ne 1$ (experimentally,
this has been studied recently for the spin glass CdCr$_{1.7}$In$_{0.3}$S$_4$
\cite{Heri02} and a colloidal glas \cite{Bell02}). 
This means that in a glass different
temperatures can be defined, depending on the way temperature is measured.
These phenomenological properties characteristic of a glass can be
partially reproduced in our simple model, even though we did {\em not}
introduce disorder into the Hamiltonian.

Specifically, we have shown, in the setting of the kinetic spherical model, 
that in the clamped case
\begin{enumerate}
\item the effective time-dependent temperature $T_{\rm eff}(t)$ as defined
in eq.~(\ref{5:Teff}) from the equal-time spin-spin correlator agrees
with the external bath temperature $T(t)$ for times larger than
$\tau_{\rm eq}$.
\item the time-evolution of two-time observables is governed by the
{\em initial} temperature $T_0$ for times up to the new time scale (if $2<d<4$)
\BEQ
\tau_{\rm eff} \sim \left( \frac{T_1 - T_c}{T_c}\right)^{-(4-d)/(2(d-2))}
\left( T_0 -T_1\right)^{-1/2}
\EEQ
distinct from both $\tau_{\rm eq}$ and $\tau$ and much larger than $\tau$ if
$T_1$ is close to criticality. Remarkably, 
$\tau_{\rm eff}=\sqrt{\tau\,\tau_{\rm dep}}$ is the geometric mean of
$\tau$ and the purely internal time scale $\tau_{\rm dep}\ne\tau$, see
(\ref{eqtaueff}). Here the temporal evolution of the 
two-time observables is delayed with respect to the evolution of the
bath temperature $T(t)$.
\item there is an intermediate regime $\tau_{\rm eq}(T_0)\ll s\ll \tau$ 
such that the system's internal degrees of freedom are in 
quasi-equilibrium while
equilibrium is not yet achieved with respect to the temporal evolution of the
bath. Then we find that the model reaches a quasi-stationary state, which
however cannot be a quasi-equilibrium state, since the fluctuation-dissipation
ratio 
\BEQ \label{gl:6:plat}
X(t,s) = \left( 1 - \frac{2\tau_{\rm eq}(T_0)}{\tau}\frac{T_0 -T_1}{T_0}
\right)^{-1}
\EEQ
takes on a value different from (albeit close to)
unity in the regime $\tau_{\rm eq}(T_0)\ll s\ll \tau$.
\end{enumerate}
In distinction to glasses, the plateau value (\ref{gl:6:plat}) is only
reached in the {\em intermediate} regime $\tau_{\rm eq}(T_0)\ll s\ll \tau$, 
before the system relaxes to equilibrium for very large 
waiting times $s\gg\tau$. 
\end{enumerate}

We stress that the clamped case arises for $\tau$ large.
At first sight, one might have naively expected that if the bath temperature
$T(t)$ is slowly changed, the system might be able to follow the evolution
of the bath through a sequence of equilibrium states. Although our explicit
results are in agreement with the adiabatic approximations they also show that
the validity of the fluctuation-dissipation theorem depends on more than the
validity of the adiabatic picture.
This observation might be of particular significance for the
interpretation of experimental studies on the time-evolution of
cooling systems (our results only hold if both the initial and final
temperatures are in the disordered phase). 
Finally, it remains an open question what
happens when the ordered phase is entered. We hope to come back
to this problem in the future.

While our results were obtained in the framework of the spherical model,
we expect that the qualitative results, in particular on when an equilibrium
state can be reached, should be of broader validity. Tests
of this expectation in other systems would be of interest.

\zeile{2}
\noindent {\large\bf Acknowledgements}\\

\noindent
We thank J-M Luck for a useful discussion. MH thanks the Centro de F\'{\i}sica
da Mat\'eria Condensada (CFMC)
of the Universidade de Lisboa for warm hospitality.

%%++++++++++++++++++++++++++++++++++++++++++++++++++++++++++++++++++++++++++++++

\end{document}